\documentclass[runningheads]{llncs}
\usepackage{hyperref}
\usepackage[T1]{fontenc}
\usepackage[utf8]{inputenc}
\usepackage{xspace}
\usepackage{graphicx}
\usepackage{listings}
\usepackage{subcaption}
\usepackage{xcolor}
\usepackage{lstautogobble}
\usepackage{color}

\definecolor{bluekeywords}{rgb}{0.13, 0.13, 1}
\definecolor{greentypes}{rgb}{0, 0.5, 0}
\definecolor{redstrings}{RGB}{171, 114, 2}
\definecolor{graynumbers}{rgb}{0.5, 0.5, 0.5}
\definecolor{goldcomments}{rgb}{0.6, 0.4, 0.08}

\lstdefinelanguage{Lola}{
  keywords=[0]{input, output, trigger, constant, import},
  keywordstyle=[0]\bfseries\color{bluekeywords},
  keywords=[1]{if, then, else, aggregate, defaults, offset, by, or, to, sin, cos, abs, hold, over, using},
  keywords=[2]{Variable, String, Int, UInt, Bool, Float32, Float64, Float, @1Hz, @5Hz, @10Hz, @100mHz, @1kHz, @1min},
  keywordstyle=[2]\color{greentypes},
    sensitive=false,
    comment=[l]{//},
    morecomment=[s]{/*}{*/},
    morestring=[b]',
    morestring=[b]"
}
\newcommand{\rtlola}[0]{RTLola\xspace}
\newcommand{\tool}[0]{\textsc{\rtlola Playground}\xspace}

\begin{document}
\title{Leveraging Static Analysis: An IDE for RTLola}
%
%
\author{Bernd Finkbeiner\orcidID{0000-0002-4280-8441} \and
  Florian Kohn\orcidID{0000-0001-9672-2398} \and
  Malte Schledjewski\orcidID{0000-0002-5221-9253}}
\authorrunning{Finkbeiner et al.}
%
\institute{CISPA Helmholtz Center for Information Security\newline66123 Saarbrücken, Germany\newline
  \email{\{finkbeiner, florian.kohn, malte.schledjewski\}@cispa.de}}
\maketitle              
\begin{abstract}
  Runtime monitoring is an essential part of guaranteeing the safety of cyber-physical systems.
  Recently, runtime monitoring frameworks based on formal specification languages gained momentum.
  These languages provide valuable abstractions for specifying the behavior of a system.
  Yet, writing specifications remains challenging as, among other things, the specifier has to keep track of the timing behavior of streams.
  This paper presents the \tool{}, a browser-based development environment for the stream-based runtime monitoring framework RTLola.
  It features new methods to explore the static analysis results of RTLola, leveraging the advantages of such a formal language to support the developer in writing and understanding specifications.
  Specifications are executed locally in the browser, plotting the resulting stream values, allowing for intuitive testing.
  Step-wise execution based on user-provided system traces enables the debugging of identified errors.
\end{abstract}
\subsubsection*{Acknowledgements.}
This work was supported by the European Research Council (ERC) Grant HYPER (No. 101055412), by DFG grant 389792660 as part of TRR 248, and by the Aviation Research Programm LuFo of the German Federal Ministry for Economic Affairs and Energy as part of “Volocopter Sicherheits-Technologie zur robusten eVTOL Flugzustands- Absicherung durch formales Monitoring” (No. 20Q1963C).
\section{Introduction}
\begin{figure}
	\centering
	\includegraphics[height=10em]{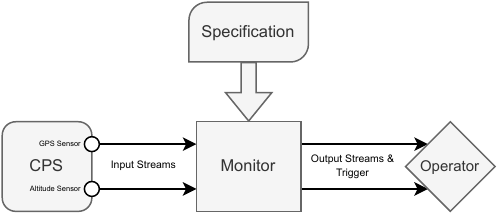}
	\caption{The stream-based Monitoring Approach}
	\label{fig:monitoring}
\end{figure}
Cyber-physical systems have become an essential part of our everyday lives.
Being safety-critical, their failure threatens humans and the environment.
Consequently, new methods are needed to ensure their correct and safe behavior.
While synthesizing or verifying such systems based on logics is an active field of research, applying these approaches to more extensive systems is computationally infeasible.
Runtime verification techniques provide scalability by monitoring the system's behavior at runtime.
This methodology has proven to be applicable in many real-world scenarios \cite{rtlola,tessla_autosar}.
In Runtime Verification, a monitoring component is deployed alongside the system, observing it and producing verdicts about its health and conformity.
Such monitors can be realized through conventional programming or generated automatically from a specification given in a formal specification language.
One class of formal specification languages adequate for such a task are stream-based specification languages.
Pioneered by Lola \cite{lola}, they process incoming data as streams from which new streams can be computed.
Trigger conditions can be defined to assess the system's state and notify an operator in case of a violation.
This stream-based monitoring approach is summarized in Figure~\ref{fig:monitoring}.
One stream-based specification language is RTLola \cite{streamlab}.
It features real-time capabilities paired with a strong type system.
Other such languages are, for example, TeSSLa \cite{tessla_playground}, and Striver \cite{striver}.

While stream-based specification languages provide useful abstractions to model the behavior of cyber-physical systems, writing correct specifications and reasoning about them is equally crucial as it is challenging \cite{lola_guarantees}.
Especially concerning autonomous aircraft, understanding the specification is essential with regard to certification and regulation conformity \cite{safe_operation}.

This paper presents the \tool\footnote{\href{https://rtlola.org/playground}{\tool: \nolinkurl{https://rtlola.org/playground}}}.
A new web-based integrated development environment for RTLola.
It eases the process from adopting runtime verification techniques to writing and testing specifications.
It is based on the \rtlola Framework extended with new static analyses that are then visualized by the tool.
For example, directed graph-based analysis results can be explored interactively, similar to Evonne \cite{evonne}, a tool for visualizing proof trees generated by automated reasoning methods.
Taking inspiration from other "playground"-style web-based IDEs for programming languages \cite{rust_playground,go_playground}, the \tool executes specifications directly in the browser based on user-provided system traces.
Monitor verdicts and intermediate values are plotted in graphs to assess the specification's correctness visually.
To ease the debugging of specifications, a method for their step-wise execution is included.
Other web-based tools for formal methods take a similar approach.
The stream-based specification language TeSSLa also features a playground \cite{tessla_playground} where users can quickly test specifications.
Yet, it does not aid the specifier in understanding static analysis results.

The rest of this paper is organized as follows:
Section~\ref{sec:examples} presents motivating examples highlighting the benefits of the \tool.
Following, Section~\ref{sec:rtlola} presents an overview of the RTLola specification language.
Section~\ref{sec:framework} gives the main points of the existing RTLola toolchain and its library structure.
In Section~\ref{sec:tool}, we present the web-based IDE for RTLola and briefly overview the tool's architecture.
Section~\ref{sec:lessons} reviews the \tool from a users perspective before 
Section~\ref{sec:conclusion} concludes the paper.
\section{Writing Specifications is Hard}
\label{sec:examples}
\begin{figure}[h]
    \centering
    \includegraphics[width=\textwidth]{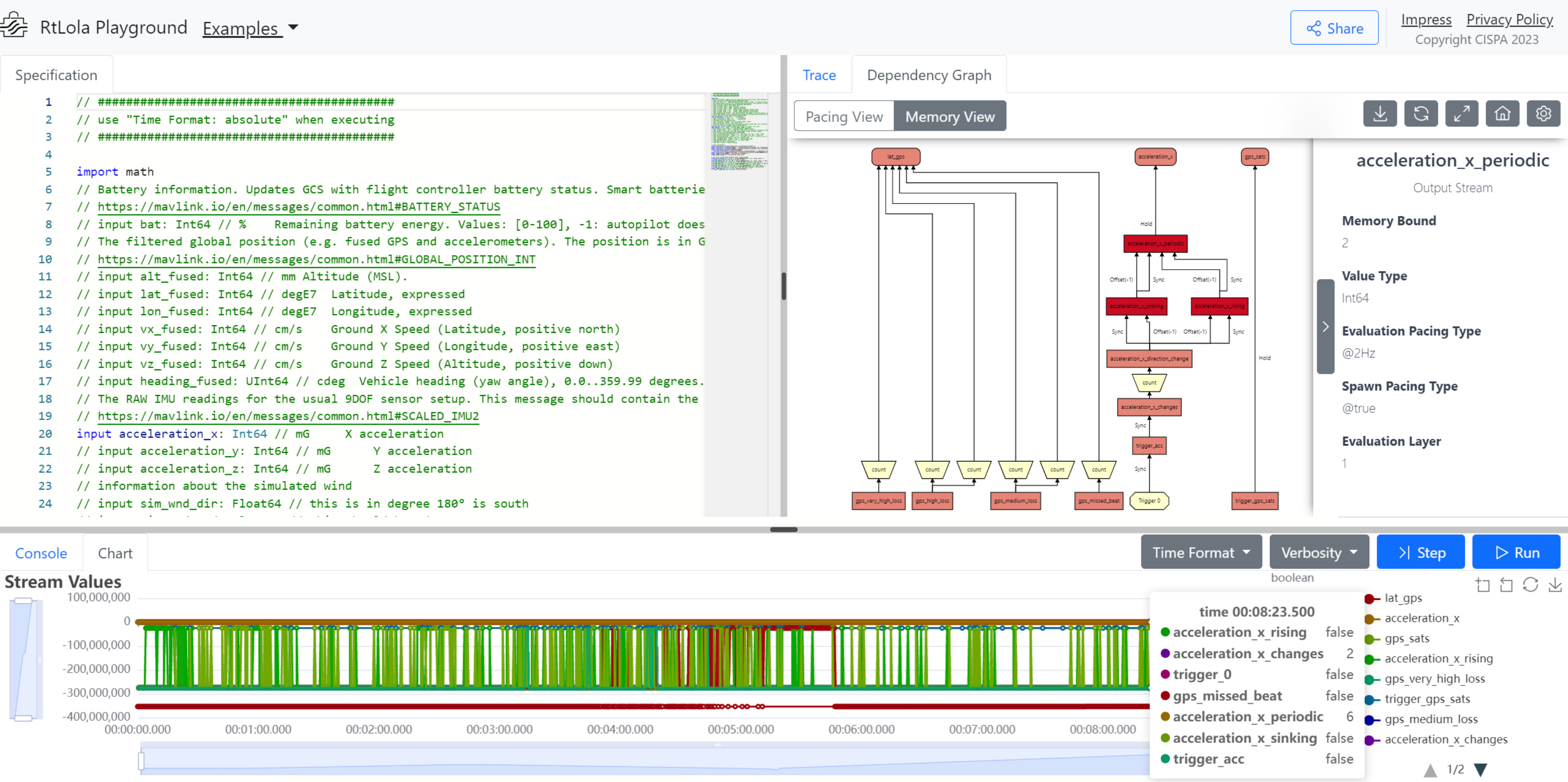}

    \caption{Screenshot of the RTLola playground. The left panel contains the specification editor, the right panel the dependency graph and the trace editor, and the bottom panel contains the output of the interpreter as the CLI output and a plot.}
    \label{fig:rtlola playground}
\end{figure}
\begin{figure}[hp]
    \centering
    \begin{subfigure}[t]{0.68\textwidth}
        \centering
        \includegraphics[width=\textwidth]{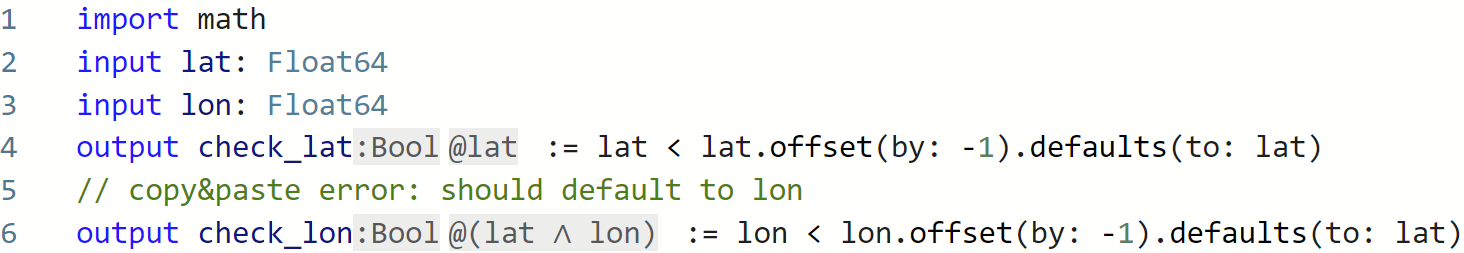}
        \caption{Specification containing an erroneous access from \emph{check\_lon} to \emph{lat}.}
        \label{fig:spec_with_additional_edge}
    \end{subfigure}
    \hfill
    \begin{subfigure}[t]{0.3\textwidth}
        \centering
        \includegraphics[width=\textwidth]{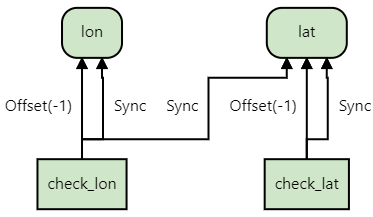}
        \caption{Dependency graph with the additional edge.}
        \label{fig:dep_graph_additional_edge}
    \end{subfigure}
    \caption{
        A specification with a typical copy\&paste error in the form of an access from \emph{check\_lon} to \emph{lat} instead of \emph{lon}.
        This error can be spotted in the editor due to a change in the inferred pacing type of the accessing stream.
        Likewise, the dependency graph shows an additional edge width breaks the symmetry and therefore can also be spotted easily.}
    \label{fig:additional edge}
\end{figure}
\noindent Writing specifications poses similar challenges as programming in general.
Large specifications do not fit into the human working memory and small errors such as simple typos or copy\&paste errors can creep in.
The \tool tackles these problems on several fronts.

As depicted in the screenshot of the user interface in Figure~\ref{fig:rtlola playground} it is divided into three sections.
The left panel features a rich text editor for specifications.
The right panel contains the editor for traces and the dependency graph, a static analysis result of the \rtlola framework.
The visualization of static analysis results is complemented by the integration of an interactive execution of the monitor on a user-provided trace.
The bottom panel features either a plot or a textual representation of the resulting stream values allowing for a quick exploration of the specification's behavior.
Just like the dependency graph, the plot also allows zooming and hiding uninteresting streams.

\rtlola is a language with a rich type system that is already used to check the specification prior to execution but showing the inferred types directly in line with the specification further improves the feedback loop.
Some of the simple copy\&paste errors such as forgetting to change the accessed stream can already have an influence on the inferred pacing type and therefore more easily spotted as seen in Figure~\ref{fig:spec_with_additional_edge}.
The \rtlola framework already uses another static analysis artifact: the dependency graph.
It consists of all streams, sliding windows, and accesses.
A more complete definition of the dependency graph is given in Definition~\ref{def:dependency graph}.
The same error as described above would lead to a different edge in the graph which can break the symmetry between copied parts as seen in Figure~\ref{fig:dep_graph_additional_edge}.

\begin{figure}[hp]
    \centering
    \begin{subfigure}[t]{0.68\textwidth}
        \centering
        \includegraphics[width=\textwidth]{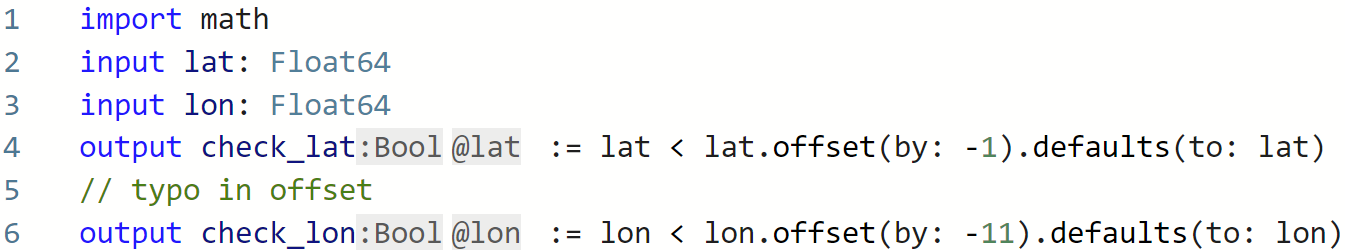}
        \caption{Specification containing an error in the offset of the access from \emph{check\_lon} to \emph{lat}.}
        \label{fig:spec with wrong offset}
    \end{subfigure}
    \hfill
    \begin{subfigure}[t]{0.3\textwidth}
        \centering
        \includegraphics[width=\textwidth]{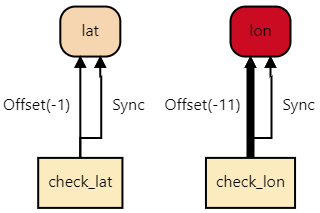}
        \caption{Dependency graph in \emph{memory view} mode.}
        \label{fig:dep graph with wrong offset}
    \end{subfigure}
    \caption{
        A specification with a typical typo in the offset of a stream access.
        This error does not change the inferred pacing type but the different offset changes the required memory of the accessed stream and therefore its color in the \emph{memory view} mode.
        In addition, the corresponding edge in the dependency graph is thicker.}
    \label{fig:wrong offset}
\end{figure}

Other simple errors such as accessing a stream with a wrong offset cannot be spotted by checking the inferred types.
A wrong offset does not change the inferred type but it changes the thickness of the edge when viewing the dependency graph in \emph{memory view} mode and potentially the buffering requirement of the accessed stream which leads to a different color as seen in Figure~\ref{fig:wrong offset}.
Similarly, a mismatch in a periodic pacing type leads to different color in the \emph{pacing view} mode.

\begin{figure}[hp]
    \centering
    \begin{subfigure}[t]{0.48\textwidth}
        \centering
        \includegraphics[width=\textwidth]{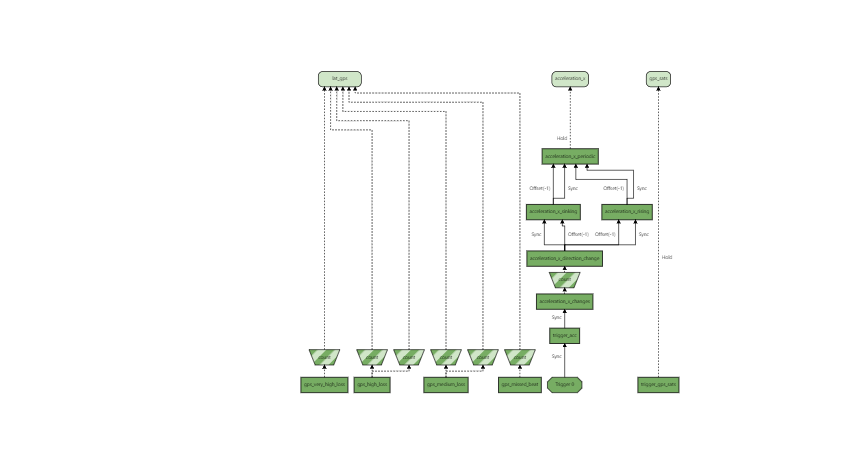}
        \caption{Fully expanded dependency graph.}
        \label{fig:dep expanded}
    \end{subfigure}
    \hfill
    \begin{subfigure}[t]{0.48\textwidth}
        \centering
        \includegraphics[width=\textwidth]{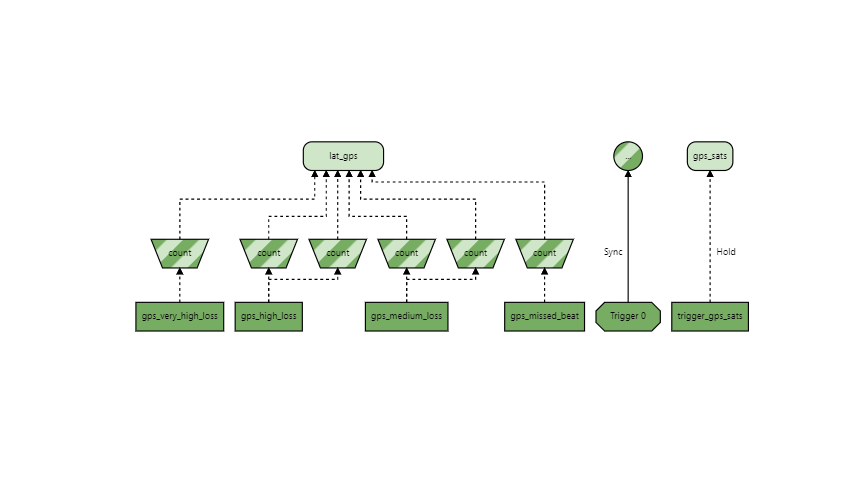}
        \caption{Dependency graph with the parts relevant only for Trigger 0 merged.}
        \label{fig:dep merged}
    \end{subfigure}
    \caption{
        The dependency graph of the same specification. Once fully expanded and once a large part of it merged to better focus on the rest.}
    \label{fig:merging}
\end{figure}

To tackle the aspect of cognitive overload, the \tool allows for merging connected nodes in the dependency graph to hide currently uninteresting parts as is demonstrated in Figure~\ref{fig:merging}.
This could be augmented on the language level by adding a module system.
Some preliminary exploration has been done in this direction.

\section{The RTLola specification Language}
\label{sec:rtlola}
In this section, we give an overview of the \rtlola specification language.
An \rtlola specification consists of input streams, representing sensor reading of the system, output streams, representing internal computations and trigger conditions, constituting an assessment of the system's health.
Furthermore, \rtlola distinguishes streams by their timing behavior.
This timing behavior is part of \rtlola's type system and is called the pacing type of a stream.
There are two disjunct timing variants:
Event-based streams are evaluated in an ad-hock manner whenever the streams they depend on produce a new value.
Periodic streams produce values at a fixed frequency.
The specification in Listing~\ref{lst:running} is used as a running example throughout this section and monitors abrupt altitude changes of an autonomous aircraft.

\begin{lstlisting}[language=lola, caption={\rtlola: A running Example.}, label={lst:running}]
input altitude : Float

output avg_altitude @1Hz := 
		altitude.aggregate(over: 1min, using: avg)
		
output altitude_diff := 
		abs(altitude - avg_altitude.hold(or: altitude))
		
trigger altitude_diff > 10.0 "Altitude changed too quickly"
\end{lstlisting}
In this specification's first line, the input stream \lstinline{altitude} is declared.
As for all input streams, only its value type, \lstinline{Float} in this case, is known.
Distinctly, \rtlola does not pose any assumptions on the timing behavior of input streams, i.e. the time when a new sensor reading arrives at the monitor remains unknown till runtime.

As the name suggests, the output stream \lstinline{avg_altitude} declared in line 3 computes the average as a sliding window over the \lstinline{altitude} input stream.
A sliding window accumulates all values of the target stream in the given time frame.
In the example above, this time frame is one minute.
Because sliding windows do not imply any timing for their evaluation, the stream has to be annotated with an explicit frequency of \lstinline{1Hz}, inducing that a new stream value, and therefore for the window, is computed every second.
Note that sliding windows must have a periodic timing to have bounded memory.
We refer the interested reader to \cite{streamlab} for more details on sliding windows.

The following output stream in line 6 computes the difference between the average altitude and the currently measured one.
It highlights an essential part of the \rtlola type system: Periodic and event-based streams must not be accessed synchronously in the same expression.
The \lstinline{altitude} access in the stream's expressions constitutes a synchronous access.
A synchronous access reads the target stream's current value and additionally binds the accessing stream's timing to the accessed stream's timing.
This guarantees that the accessing stream is only evaluated if the accessed value exists.
As events can never be assumed to happen with a fixed frequency the type-checking procedure fails if a stream accesses both a periodic and an event-based stream synchronously, as no common timing can be determined in which both accessed values are always guaranteed to exist.

To resolve this, the stream in line 6 of Listing~\ref{lst:running} uses a \lstinline{hold} access to the timed average stream.
A \lstinline{hold} access refers to the last computed value of a stream.
If no such value exists, a provided default value is substituted.
Last, a \lstinline{trigger} is defined to alert the operator if the current altitude deviates more than ten units from its average.
\section{The RTLola Framework}
\label{sec:framework}
The \rtlola framework is split into two purviews.
The \rtlola Frontend is responsible for parsing and analyzing specifications.
A Backend handles the event input, executes the specification, and forwards the output to the user.
Executing a specification can follow different paradigms.
The specification can be interpreted by the \rtlola Interpreter or cross-compiled to a programming or hardware description language by the \rtlola Compiler.

\begin{figure}
	\centering
	\includegraphics[width=\linewidth]{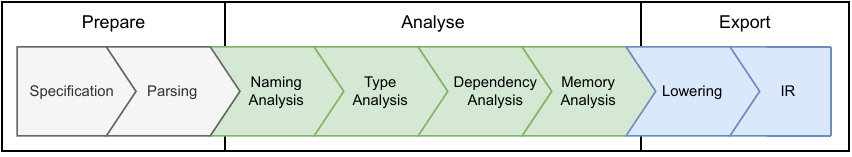}
	\caption{An overview of the RTLola Frontend}
	\label{fig:frontend}
\end{figure}

\subsection{The \rtlola Frontend}
The \rtlola Frontend\footnote{\href{https://crates.io/crates/rtlola-frontend}{\rtlola Frontend: \nolinkurl{https://crates.io/crates/rtlola-frontend}}} is divided into multiple phases consisting of multiple stages.
Figure~\ref{fig:frontend} depicts an overview of these stages.
In the first phase, the specification is parsed into an abstract syntax tree.
The AST is then transformed into its de-sugarized form by representing all syntactic sugar constructs by basic \rtlola expressions.

Next, the \textbf{naming analysis} checks the AST for duplicated or undefined stream names.
For example, this stage rejects all specifications in which the same stream is defined multiple times.
Afterward, the AST is transformed into a high-level intermediate representation by replacing stream name occurrences with numerical ids based on the previous analysis.

The \textbf{type analysis} infers a value and a pacing type for every stream.
The value type of a stream determines the semantics of produced values.
The value type system is similar to the one of programming languages.
Consequently, \rtlola also supports the usual value types, such as signed and unsigned integers, floats, strings, and booleans, including combinations of those types through tuples.

The pacing type of a stream determines the temporal behavior of a stream, i.e., when a new stream value is computed.
As described in Section~\ref{sec:rtlola}, there are two classes of pacing types.
An event-based type (e.g. \lstinline{@(lat $\land$ lon)} in Figure~\ref{fig:spec_with_additional_edge}) signals that the stream is computed whenever an event occurs.
An event is a combination of input streams receiving a new value synchronously.
Such a combination is described through a positive boolean formula over input streams.
A synchronous access from one event-based stream to another is allowed iff the accessing stream's pacing type implies the pacing of the accessed stream.

A periodic type (\lstinline{@1Hz}) indicates that a stream is computed at a fixed frequency.
A synchronous access from one periodic stream to another is allowed iff the accessing stream's frequency divides the frequency of the accessed stream.
A synchronous access from an event-based stream to a periodic stream or vice versa is not allowed.

The \textbf{dependency analysis} computes the dependency graph of the specification and checks its well-formedness as presented in \cite{lola}.
The dependency graph of a specification is defined as follows:
\begin{definition}
\label{def:dependency graph}
	The dependency graph of a specification with inputs $i_1,...,i_m$ and outputs $o_1, ..,o_n$ is a directed weighted multi-graph $G = \langle V, E \rangle$ with $V=\{i_1,...,i_m,o_1,...,o_n\}$. An edge $e=\langle o_i, o_k, w \rangle$ is in $E$ iff the expression of $o_i$ contains $o_k.\textit{offset}(\textit{by: }w,\textit{ or: } c)$ as a sub-expression or $e=\langle o_i, i_k, w \rangle$ if $i_k.\textit{offset}(\textit{by: }w,\textit{ or: } c)$ is a sub-expression. Synchronous accesses are reflected as offsets by 0.
\end{definition}
To recap, the dependency graph of a not well-formed specification contains a cycle with an accumulated weight of 0.
As a result, specifications such as:
\begin{lstlisting}
	output a:= b
	output b:= a
\end{lstlisting}
are rejected.

The \textbf{memory analysis} computes a per stream upper bound for the number of values that must be stored as defined in \cite{streamlab}.
Intuitively, if the maximal offset a stream is accessed with is 2, then three values must be stored for that stream, including the current value.

After the high-level intermediate representation is validated through the static analyses, it is lowered into the final intermediate representation, dropping information irrelevant to backends.

\begin{figure}
	\centering
	\includegraphics[height=10em]{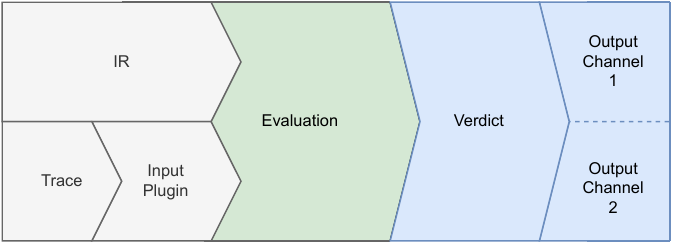}
	\caption{An overview of the RTLola Interpreter}
	\label{fig:interpreter}
\end{figure}

\subsection{The RTLola Interpreter}
The \rtlola Interpreter\footnote{\href{https://crates.io/crates/rtlola-interpreter}{\rtlola Interpreter: \nolinkurl{https://crates.io/crates/rtlola-interpreter}}} is an interpreter for RTLola specifications.
Developed for the rapid prototyping of specifications, it forms the basis for evaluating specifications in the \tool.
Figure~\ref{fig:interpreter} shows an overview of the interpreter architecture.
The specification can be processed directly in the form of its intermediate representation.
To handle a variety of trace formats, the interpreter adds a layer of indirection through input plugins that translate events from their trace representation to an internal representation.
This way, the interpreter can accept events in various formats like CSV, network packet capture (PCAP), or serialized as bytes.

Each event starts a new evaluation cycle in which all periodic streams up to the current point are evaluated before all event-based streams are evaluated that were activated by the event.
Which information the produced verdict contains is up to configuration and ranges from trigger messages to the current state of all streams.
The verdict is forwarded to one or multiple output channels responsible for displaying or forwarding that information.
\section{Tool Overview}
\label{sec:tool}

\begin{figure}
  \centering
  \includegraphics[width=0.8\textwidth]{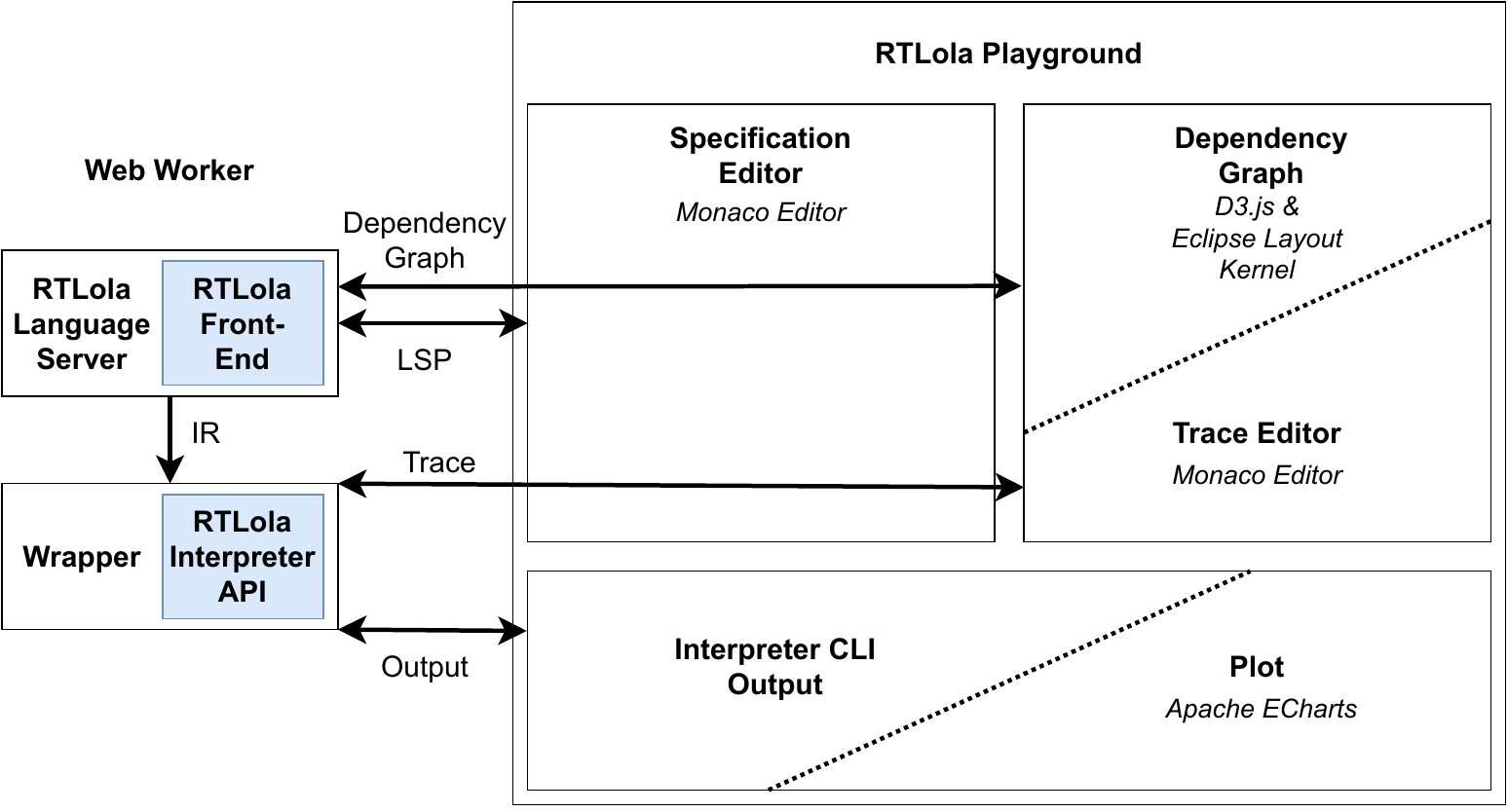}

  \caption{Simplified overview of the main software components and their interaction. The blue boxes are deployed as WebAssembly compiled from Rust code. The other parts are implemented in TypeScript. Web workers from third-party libraries are not shown.}
  \label{fig:architecture}
\end{figure}

The \tool is a progressive web application based on the \emph{Vue}~\cite{vue} framework and in general written in \emph{TypeScript}~\cite{typescript}.
An overview of the main components and their interaction is shown in Figure~\ref{fig:architecture}.
The components communicate mostly via shared \emph{Pinia}~\cite{pinia} stores.

The \rtlola framework is implemented in the \emph{Rust}~\cite{rust} language.
This allows for easy compilation to \emph{WebAssembly}~\cite{WebAssembly} which means, that the code running in the browser matches the code powering the \rtlola Interpreter executable.

The text editing is provided by the \emph{Monaco Editor}~\cite{monaco} which is extracted from \emph{Visual Studio Code}~\cite{vscode}.
For the specification editor, we also provide a language server for \rtlola which is mostly a wrapper written in TypeScript and Rust around the \rtlola Frontend.
This ensures that the user gets the same errors as if they were using the \rtlola interpreter executable.
The language server mostly communicates with the specification editor via the \emph{Language Server Protocol}~\cite{lsp} to enable inline hints and diagnostics but it also provides additional artifacts such as the dependency graph and the intermediate representation of the specification.

The dependency graph is mostly based on the \emph{D3.js}~\cite{d3} library while the layout is handled by the \emph{Eclipse Layout Kernel (elkjs)}~\cite{elkjs} guided by information from the static analysis.
A thin wrapper written in TypeScript and Rust around the \rtlola interpreter API allows for executing monitors directly in the browser.
One can inspect the CLI output as if one were to use \rtlola interpreter executable but in addition the playground also contains a plot of all scalar numerical and boolean stream values.
The plot is based on \emph{Apache ECharts}~\cite{echarts} which provides the typical interactions such as hover, filtering, and zooming.

\section{Application Scenarios}
\label{sec:lessons}
This section reviews the benefits of the \tool by considering two usage scenarios.

Firstly, new users of \rtlola can quickly test their mental model about stream-based specification languages.
They can run specifications and try them against different traces without interacting with a complicated command line interface or dealing with an installation process.
Additionally, we plan to integrate an interactive tutorial directly into the \tool to lower the entry barrier further.
There have been many studies on how and when to give feedback during learning~\cite{doi:10.3102/0034654307313795,doi:10.3102/0034654314564881}.
More elaborate feedback than simple right/wrong improves learning and in the case of the \tool{} we believe that for type errors, showing the expected and the actual type strikes a good balance between enough information and feedback complexity.
For learning basic programming skills immediate feedback seems to work best for beginners.
Immediate feedback can be detrimental if it leads to simply gaming the system until a correct answer is found but as we do not have a given task this is not the case for the \tool{}.

Secondly, expert users of \rtlola can use the \tool to get better insights into the specification's memory consumption, timing behavior, or locality.
Exemplary, expert users can use the dependency graph to identify possible optimizations.
In Figure~\ref{fig:merging}, one can see that the \lstinline{count} aggregation is repeated five times with an identical duration.
This can be optimized by outsourcing this aggregation into a separate stream.

These application scenarios show, that the \tool is not only suitable for users of different knowledge backgrounds but can also be a step towards a wide adoption of runtime verification techniques.

\section{Conclusion}
\label{sec:conclusion}
This paper presented the \tool, a web-based integrated development environment for the stream-based specification language \rtlola.
Built with cutting-edge web technologies like WebAssembly and web workers, it features a rich text editor for specifications, integrated testing and debugging capabilities, and interactive visualizations for static analysis results.
We have demonstrated how specific specification errors can be identified using either the editor's feedback or the static analysis results.
We elaborated on how different user groups can use the playground to their advantage and hence conclude that the \tool helps specifiers to write correct specifications faster while keeping the entry barrier for new users low.

In the future, we plan to reuse most of the components of the tool in an extension for the Visual Studio Code editor and integrate an interactive tutorial into the playground.

Lastly, we encourage other community members to port their research tools to the browser.
Many modern compiler toolchains support a compilation to Web\-Assembly, which keeps the overhead feasible.
A web-based tool enables easy adoption and makes research easier to reproduce and transfer.

%
\bibliographystyle{splncs04}
\bibliography{references}
\end{document}